\documentclass{article}[12pt,a4paper]
\usepackage{graphicx}  
\usepackage{dcolumn}   
\usepackage{bm}        
\usepackage{amssymb}   
\usepackage[]{amsmath,amssymb}
\usepackage{graphics,epsfig}
\usepackage[latin1]{inputenc}
\usepackage[T1]{fontenc}
\usepackage{amsfonts}
\usepackage{latexsym}
\usepackage[english]{babel}
\newcommand{\be}{\begin{equation}}
\newcommand{\ee}{\end{equation}}
\newcommand{\beq}{\begin{equation}}
\newcommand{\eeq}{\end{equation}}
\newcommand{\bea}{\begin{eqnarray}}
\newcommand{\eea}{\end{eqnarray}}

\def\be{\begin{equation}}
\def\ee{\end{equation}}
\def\ba{\begin{eqnarray}}
\def\ea{\end{eqnarray}}

\begin{document}


\title{Entanglement entropy as a witness of the Aharonov-Bohm effect in QFT}
\author{Ra\'{u}l E. Arias, David D. Blanco, Horacio Casini\\
{\sl Centro At\'omico Bariloche,
8400-S.C. de Bariloche, R\'{\i}o Negro, Argentina}}

\maketitle

\begin{abstract}
We study the dependence of the entanglement entropy with a magnetic flux, and show that the former quantity witnesses an Aharonov Bohm-like effect. In particular, we consider free charged scalar and Dirac fields living on a two dimensional cylinder and study how the entanglement entropy for a strip-like region on the surface of the cylinder is affected by a magnetic field enclosed by it.
\end{abstract}

{\sl \bf Introduction.---} The Aharonov-Bohm effect (AB) is a fundamental quantum phenomenon in which an electrically charged particle is affected by an electromagnetic potential $A_{\mu}$, even if the magnetic and electric components of this field vanish in the region where the particle is confined. The AB effect emerges as a consequence of the fact that the circulation of $A_{\mu}$ around a curve $C$ ($\Phi:=\oint_C A_{\mu}dx^{\mu}$) can be sensed in the wave function of a charged particle $\psi\left(x\right)$, which acquires an additional phase factor $e^{i e\Phi}\psi\left(x\right)$, regardless of the precise values $A_{\mu}$ takes on the region where the particle is confined. This phase factor can therefore be seen in interference experiments of particles traveling in different paths.

This effect has first been noted by W. Ehrenberg and R. Siday in 1949 \cite{siday} and Y. Aharonov and D. Bohm in 1959 \cite{bohm}, and has been observed in the laboratory \cite{exp}.
The response of the expectation value of certain QFT operators (on cylindrical geometries) under a magnetic flux were recently analyzed using holography on \cite{Montull}. In the condensed matter literature, it has been also studied the effect of the magnetic flux on unconventional superconductors with cylindrical geometry (see for example \cite{cmt}).  In this work, we analyze the AB effect on the vacuum fluctuations using entanglement entropy.

The entanglement entropy refers to the von Neumann entropy $S\left(V\right)$ of the vacuum state reduced to a region $V$ of the space
\be
S\left(V\right)=-tr \left(\rho_V \log\rho_V\right)\,;\label{def}
\ee
with $\rho_V$ the reduced density matrix. It essentially measures the entropy contained in the vacuum fluctuations in $V$.

Though it originated in an attempt to explain the entropy of black holes, the entanglement entropy has nowadays become an exceptional theoretical tool that provides new insights into a variety of topics in physics. In condensed matter theory, it can be used to distinguish new topological phases or different critical points \cite{wenx,cardy0}. It has also been proposed as a useful probe of phase transitions in gauge quantum field theories \cite{igor0} and has brought a new perception on the structure of renormalization group flows \cite{twoD,pcon,cthem}, being essential to prove the c-theorem in three dimensions \cite{threeD}.

In this paper we show that the entanglement entropy exhibits a dependance on the Aharonov-Bohm phase $\Phi$, thus becoming an attractive tool to explore related topological phenomena. Specifically, we compute the entanglement entropy for free charged scalar and Dirac fields in the presence of an electromagnetic potential in a simple two dimensional example.

\begin{figure}
\centering
\leavevmode
\epsfysize=4cm
\epsfbox{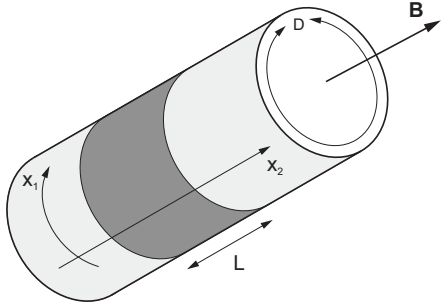}
\bigskip
\caption{The geometry used in the text. Coordinate $x^1$ is compactified in a circle of size $D$ and we look at the entropy of an annular strip of width $L$. The external gauge field induces a phase in the $x^1$ direction. }
\label{figu1}
\end{figure}

{\sl \bf The AB effect on entanglement entropy.---} We are going to analyze the case of a free scalar field, $\phi$, charged with respect to an external gauge field, $A_\mu$, which is pure gauge in the region of interest. Hence we start with the Lagrangian
\be
{\cal{L}}=-(\partial_\mu+i e A_\mu)\phi^* (\partial^\mu-i e A^\mu)\phi-m^2 \phi^* \phi\label{uno}
\ee
for a charged free scalar field with mass $m$. To keep the calculation as simple as possible we consider the case of a space  compactified in a circle of size $D$ in the $x^1$ direction (see figure \ref{figu1}) with periodic boundary conditions for the field, $\phi(x^0,0,x^2,...,x^{d})=\phi(x^0,D,x^2,...,x^{d})$. We choose a constant gauge field in the $x^1$ direction. When the pure gauge field $A_\mu=\partial_\mu \alpha(x)$ is turned on we can eliminate it by a gauge transformation
\be
\phi(x)\rightarrow  e^{-i e \int_{\tilde{x}}^x dy^\mu A_\mu(y)}\,\phi(x)\,,
\ee
where the base point $\tilde{x}$ of the integral is arbitrary. This has the consequence that the scalar field has now the following boundary condition
\be
\phi(x^0,0,...,x^{d})=e^{-i e \oint A_1 dx^1 }\phi(x^0,D,...,x^{d})\,.
\ee
The integral
\be
e \oint A_1 dx^1=\varphi=e \Phi
\ee
can be thought as proportional to the flux $\Phi$ of a magnetic field through the circle $S^1$. This magnetic field is fully outside of the space, and its effect on the scalar field is only through the AB effect. It gives a phase $\varphi$ on the boundary condition of the field $\phi$ which is now decoupled from any external sources.

We can use Lagrangian (\ref{uno}) with $e A_1=\varphi/D$ constant, and decompose it into Fourier modes  in the $x^1$ direction, $\phi=\sum_{n} e^{-i 2 \pi n x/D} \phi^{(n)}$, 
\be
{\cal{L}}=\sum_{n=-\infty}^\infty  \left(-\partial_\mu \phi^{(n)*} \partial^\mu \phi^{(n)}-\left(m^2+ \frac{(2\pi n+\varphi)^2}{D^2}\right) \phi^{(n)*} \phi^{(n)}\right)\,,
\ee
where now the space has $d-1$ dimensions with coordinates $x^2,...,x^{d}$.

The entropy of a strip of width $L$ around the cylinder (see figure \ref{figu1}) will be given by the sum over the different modes of the entropies of these massive fields living in dimension $d-1$. For two spacial dimensions $d=2$, as long as we are interested in evaluating the entropies of annulus around the direction $x^1$, the calculation is equivalent to the entropy of an interval of size $L$ in one dimension, for an infinite tower of massive fields with masses given by
\be
M(n,\varphi)=\sqrt{m^2+ \frac{(2\pi n+\varphi)^2}{D^2}}\,.\label{masses}
\ee

The same dimensional reduction holds for free Dirac fields, where the effective masses for the $d=2$ fields are given again by (\ref{masses}). Hence, the entanglement entropy of the annulus is
\be
S(L,m,\phi)=\sum_n S_1(L,M(n,\varphi))\,,\label{sum}
\ee
where $S_1(L,M)$ is the vacuum entropy for the massive $d=1$ field of mass $M$.

This is given by
\be
S_1(L,M)=-\int_{L M}^\infty dy \, \frac{C(y)}{y} -C(0) \log(M \epsilon)\,.\label{pp}
\ee
Here $\epsilon$ is a short distance ultraviolet cutoff, $M$ stands for the effective mass of the field, and
\be
C(ML)=L \frac{dS(M,L)}{dL}
\ee
is the entropic C-function \cite{twoD}. This is positive and monotonically decreasing. For zero $ML$ it takes the value $C(0)$ given by one third of the conformal central charge in the limit $M\rightarrow 0$. This is $C(0)=1/3$ for Dirac fermions and $C(0)=2/3$ for a complex scalar. For large mass $C(ML)$ is exponentially decreasing. More precisely, the limits of small and large argument for this function are \cite{review}
\be
C(y)\simeq \frac{2}{3}+\frac{1}{\log(y)}+...  \hspace{.8cm}\textrm{for}\,\, y\ll 1\,, \hspace{1.5cm} C(y)\simeq \frac{1}{2}\, y \,K_1(2 y) \hspace{.8cm}\textrm{for}\,\, y\gg 1\,,\label{dixi}
\ee
for a complex scalar, and
\be
C(y)\simeq\frac{1}{3}-\frac{1}{3}\,y^2 \,\log^2(y)+...  \hspace{.8cm}\textrm{for}\,\, y\ll 1\,, \hspace{1.5cm} C(y)\simeq \frac{1}{2}\, y\, K_1(2 y) \hspace{.8cm}\textrm{for}\,\, y\gg 1\,,\label{dixit}
\ee
for a Dirac field. The expressions for short distances are the leading logarithmic terms. The complete C-function can be calculated numerically with high precision by integrating the solutions of an ordinary differential equation \cite{review}.

The first term in (\ref{pp}) gives the shape of the one dimensional entropy as a function of $L$. We have to include the second term in (\ref{pp}) which only depends on the mass. This gives the dependence on mass of the entropy saturation constant for large $L$ \cite{cardy0}, and will be affected by changes on the mass due to the magnetic flux, eq. (\ref{masses}). The cutoff dependence in (\ref{pp}) does not play a role because we want evaluate how the entropy changes with the magnetic flux. It gives a constant overall ambiguity which is independent on the mass and $L$.  For a scalar field the entropy includes an additional $L$ independent term that depends on the mass
\be
\log(\log(- M \epsilon))\,.\label{vuelta}
\ee
This is due to infrared divergences for massless scalars in two dimensions \cite{review}. However, this mass dependent term have to be thought as giving an overall infrared constant term because its derivatives with respect to mass vanish for the limit of small cutoff. Then, we are neglecting this term in the following.

We can focus on the universal part of the change of entropy with magnetic flux by computing the quantity
\be
S(\varphi)=\int_0^\varphi d\varphi^\prime\frac{d}{d\varphi^\prime} S(L,m,\varphi^\prime) =S(L,m,\varphi)-S(L,m,\varphi=0)\,.
\ee

The contribution to $S(\varphi)$ of the second term in (\ref{pp}) is given by
\bea
&&-\int_0^\varphi d\varphi^\prime \sum_{n=-\infty}^\infty C(0)\,\,\frac{2\pi n+\varphi^\prime}{m^2 D^2+(2 \pi n+\varphi^\prime)^2}=-
\int_0^\varphi d\varphi^\prime \frac{C(0)\,\sin{\varphi^\prime}}{2 (\cosh(m D)-\cos(\varphi^\prime))}\nonumber\\
&&\hspace{8.cm}=-\frac{C(0)}{2}\log\left(\frac{\cosh(m D)-\cos(\varphi)}{\cosh(m D)-1}\right)\,.
\eea
 This is independent of the width of the strip $L$, and is always negative.

 \begin{figure}[t]
\centering
\leavevmode
\epsfysize=5cm
\epsfbox{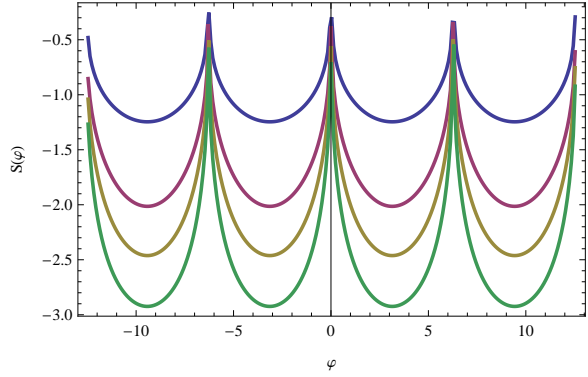}
\bigskip
\caption{$S(\varphi)$ for the massless scalar and various ratios of $L/D$. From top to bottom: $L/D=1/10, 1/2, 1, 2$. We have set the infrared divergent constant $\gamma=2$ in this picture. Notice the shape of the fluctuations do not tend to vanish for small
 $L/D$, but slowly get flatter as $L/D\rightarrow 0$.}
\label{figu2}
\end{figure}

Hence, setting $S(\varphi=0)=0$ in this way we have
\be
S(\varphi)=-\sum_{n=-\infty}^\infty  \int_{L M(n,\varphi)}^\infty dy \, \frac{C(y)}{y}
+\sum_{n=-\infty}^\infty  \int_{L M(n,0)}^\infty dy \, \frac{C(y)}{y}
-\frac{C(0)}{2}\log\left(\frac{\cosh(m D)-\cos(\varphi)}{\cosh(m D)-1}\right)\label{complete}\,.
\ee
This expression is finite, showing the $\varphi$ dependent term is regularization independent.
  Some general features of $S(\varphi)$ follow directly from (\ref{complete}) without further calculation.
Evidently, from (\ref{complete}) the entropy $S(\varphi)$ will be a periodic function of the phase $\varphi$ with period $2 \pi$. When an integer number of quantum flux $\Phi=e/2\pi$ runs through the cylinder we have $S(\varphi)=0$ and there is no net effect on the vacuum entropy. From (\ref{masses}) we can also see the effect is symmetrical under $\varphi\rightarrow -\varphi$ and $\varphi\rightarrow \pi-\varphi$. We can compute $S(\varphi)$ numerically from the knowledge of the C-function. The result shows $S(\varphi)$ is always negative; the maximum of $|S(\varphi)|$ is achieved for $\varphi=\pi$. This means the AB effect always decreases the entanglement with respect to the vacuum without magnetic field.

{\sl \bf Various limits.---} In order to study the massless case we begin by considering equation (\ref{complete}) for $m D\ll 1$ and $mL\ll 1$. Up to first order in $m D$ the third term in (\ref{complete}) gives
\be
C(0)\log\left(m D\right)-\frac{C(0)}{2}\log\left(2-2\cos\left(\varphi\right)\right)\,.\label{loga}
\ee
We can set $m=0$ in the first infinite summation of (\ref{complete}) and no divergences will arise (unless $\varphi$ is an integer multiple of $2\pi$). The second sum carries a divergence for the mode $n=0$ when we take $m=0$, but we can easily verify that it cancels out with the logarithmic term given by (\ref{loga}). We isolate the term with $n=0$ in this second summation and extract the logarithmic term for $mL\ll 1$
\be
\int_{m L}^{\infty}dy\frac{C\left(y\right)}{y}\simeq -C(0)\log\left(m L\right)+\gamma  \,,\label{tutu}
\ee
where
\be
\gamma=\lim_{y_0 \rightarrow 0}\left(\int_{y_0}^{\infty}dy\frac{C\left(y\right)}{y}+C(0) \log(y_0)\right)\,.
\ee
For a Dirac field we have
$\gamma \simeq-0.528$. On the other hand, for a scalar field, $\gamma$ is controled by infrared physics and can be large. If the infrared cutoff for the zero mode is set by a small mass we have $\gamma\sim -\log(-\log(m L))$. This is due to the first subleading term in the small $ML$ expansion of the $C$ function, eq. (\ref{dixi}). If some other mechanism set the infrared cutoff this can greatly change. For example, imposing an antiperiodic boundary condition in the $x^2$ direction we would have $\gamma\sim -\log(R/L)$, with $R$ the compact size of the $x^2$ direction.

\begin{figure}[t]
\centering
\leavevmode
\epsfysize=5cm
\epsfbox{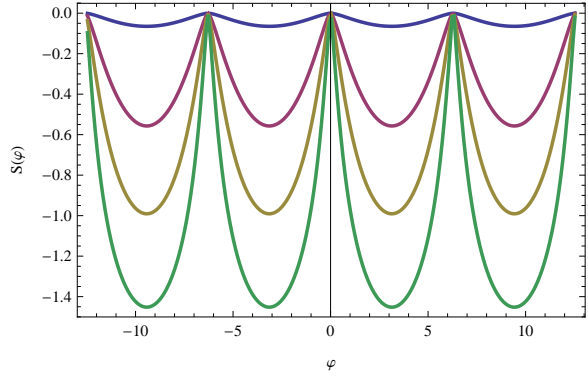}
\bigskip
\caption{$S(\varphi)$ for the massless Dirac field and various ratios of $L/D$. From top to bottom: $L/D=1/10, 1/2, 1, 2$. For large $L/D$ the curves are similar in shape, but half the overall size, as the ones for the scalar (compare with figure \ref{figu2} and formula (\ref{contri})). For small $L/D$, contrarily to the scalar case, the function $S(\varphi)$ decays to zero.  }
\label{figu3}
\end{figure}

If we write the complete expression for $S\left(\varphi\right)$ the logarithmic terms involving $m$ in (\ref{loga}) and (\ref{tutu}) cancel out and we get the expression for the entropy of the massless field
\be
S(\varphi)=\sum_{n\neq 0} \int^{\left|\frac{\left(2\pi n+\varphi\right)L}{D}\right|}_{\left|\frac{2\pi n L}{D}\right|} dy \, \frac{C(y)}{y}-\int_{\left|\frac{L\varphi}{D}\right|}^{\infty} dy\, \frac{C(y)}{y}-\frac{C(0)}{2}\log\left(2-2\cos\left(\varphi\right)\right)+\gamma-C(0)\log\left(\frac{L}{D}\right)\,.
\ee
Naturally this is a function of $L/D$. Figures (\ref{figu2}) and (\ref{figu3}) show $S(\varphi)$ for some values of $L/D$ for the scalar and Dirac fields respectively.

When $L/D\gg 1$ and $|L/D\,\varphi|\gg 1$ the first two terms are exponentially small (taking $\varphi\in (-\pi,\pi)$), and the shape of the oscillations is given by
\be
S(\varphi)=-\frac{C(0)}{2}\log\left(2-2\cos\left(\varphi\right)\right)+\gamma-C(0)\log\left(\frac{L}{D}\right)\,.\label{contri}
\ee
Excepting for a factor of two and an overall additive constant, this is the same for fermions and scalars.
The maximal size $|S(\pi)|$ of the oscillations is in this case
\be
|S(\pi)|=\frac{C(0)}{2}\log(4)-\gamma+C(0)\log\left(\frac{L}{D}\right)\,.
\ee
This can be as large as we want for large $L$ and fixed $D$. The reason for this large variations is that the one dimensional massless $n=0$ mode has an entropy increasing logarithmically with $L$, and this is cutoff by the effective mass provided by the magnetic field. However, it is interesting to note that the dependence on $\varphi$ in (\ref{contri}) includes a coherent contribution  of all modes through the term $-C(0) \log(M)$ in the entropy, even if for $L/D\gg 1$ these have large masses.

The effect of the inessential infrared divergence (\ref{vuelta}) reappears in $S(\varphi)$ for the massless scalar field through the infrared divergent constant $\gamma$, and the large variations of the entropy for $\varphi\rightarrow 0$. The change of entropy for a massless scalar field with and without magnetic field is infrared divergent for any non zero value of the flux. This large susceptibility is not present for the fermion fields because it is due to the classical zero mode of the scalar field. However the large value of $\gamma$ does not affect finite variations of $S(\varphi)$ for different $\varphi\neq 0$. It does not change the shape of the curves away from $\varphi=0$ but just displaced them to large negative values (see figure \ref{figu2}).

For small $L/D\ll 1$ the different modes add incoherently and this cuts the size of the oscillations. In this limit of small width of the annulus is better to study directly the derivative $S^\prime(\varphi)$. According to (\ref{masses}) and (\ref{sum}) in the massless limit this is
\be
S^\prime(\varphi)=\frac{L}{D} \sum_{n=-\infty}^\infty f\left(\frac{L}{D} (2 \pi n+\varphi )\right)\,, \label{sss}
\ee
where
\be
f(x)=\frac{C(|x|)-C(0)}{x}\,.
\ee
The function $f(x)$ is antisymmetric and falls to zero exponentially at infinity. If $f(x)$ was analytic we could use Euler MacLaurin formula in (\ref{sss}) to conclude that $S^\prime(\varphi)$, and hence $S(\varphi)$, vanish exponentially fast with $L/D$ for small $L/D$. However, this is not the case since $f(x)$ is non analytic at the origin, going as $f(x)\sim -1/3 x \log(|x|)^2$ for fermions and $f(x)\sim(x \log(|x|))^{-1}$ for scalars (see (\ref{dixi}), (\ref{dixit})). As a consequence, the amplitude of the oscillations falls as $(L/D)^2 \log(L/D)$ for the fermions in the limit of small $L/D$, while the derivative $S^\prime(\varphi)$ falls only logarithmically, as $(\log(L/D))^{-1}$, for the scalar field (where $\varphi$ is held fixed as $L/D\rightarrow 0$). This difference can be appreciated in the figures \ref{figu2} and \ref{figu3}.

\begin{figure}
\centering
\leavevmode
\epsfysize=5cm
\epsfbox{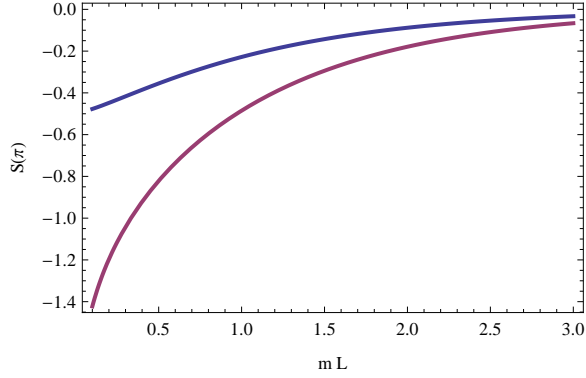}
\bigskip
\caption{$S(\pi)$ (giving the maximal size of the change of entropy) as a function of mass for $L/D=1$. For large mass $S(\pi)$ goes to zero exponentially fast. For $m\rightarrow 0$ it tends to a limit for the fermion (curve at the top) and it gets to $-\infty$ for the scalar (curve at the bottom). }
\label{figu4}
\end{figure}

For massive fields the effect of the magnetic field on the entropy is reduced with respect to the massless case. For large $m L\gg 1$ the first term of (\ref{complete}) gives an exponentially small number
$\sim 1/2 (m L) K_1(2 m L)\sim (mL)^{1/2} e^{-2 m L}$.
In this situation the third term of (\ref{complete}) gives the leading part of the entropy
\be
S(\varphi)\sim -\frac{C(0)}{2}\log\left(\frac{\cosh(m D)-\cos(\varphi)}{\cosh(m D)-1}\right)\,.
\ee
When we also have $m D\gg 1$, this last term gives again an exponentially small number
\be
-C(0)\,e^{-m D}(1-\cos\left(\varphi\right))\,,
\ee
this time with a pure sinusoidal form. We show the lowest value of the entropy $S(\pi)$ for a particular value $L/D=1$ as a function of the mass in figure \ref{figu4}.

{\sl \bf AB effect on mutual information.---} Instead of considering the entanglement entropy of an annulus of size $L$ we could as well think in the mutual information $I(A,B)=S(A)+S(B)-S(A\cup B)$ between two semi-infinite half-cylinders $A$ and $B$ separated by a distance $L$ in the $x^2$ direction. This has the advantage of being regularization independent from the beginning. The calculation proceeds by dimensional reduction in the same way as for the entropy but we have to use the one dimensional mutual information $I_1(ML)$ for a field of mass $M$ between two half-lines in one dimension separated by $L$. This gives
\be
I(L,m,\varphi)=\sum_n I_1(L \sqrt{m^2+(2 n\pi+\varphi)^2} )\,.
\ee
Taking into account that $dI_1(LM)/dL=-dS_1(A\cup B)/dL$, $I_1(ML)$ vanish for large $L$, and that $S_1(A\cup B)$ is equal to the entropy of its complement $S_1(L)$, we have
\be
I_1(L M)=\int_{L M}^\infty dy \, \frac{C(y)}{y}\,.
\ee
This is just the opposite of the entropy (\ref{pp}), but it does not contain the boundary term $\log(M)$ which is independent of $L$ and cancel in the mutual information. Then we get for the variation of the mutual information a formula similar to (\ref{complete}) but without the last term,
\be
I(\varphi)=I(L,m,\varphi)-I(L,m,0)=\sum_{n=-\infty}^\infty  \int_{L M(n,\varphi)}^\infty dy \, \frac{C(y)}{y}
-\sum_{n=-\infty}^\infty  \int_{L M(n,0)}^\infty dy \, \frac{C(y)}{y}\,.\label{nunu}
\ee
Concavity of the one dimensional entropy gives $I^{\prime\prime}_1(ML)>0$, and this in turn implies that the sum in (\ref{nunu}) is decreasing for $\varphi\in(0,\pi)$, and in consequence $I(\varphi)$ is always negative, achieving its minimum for $\varphi=\pi$. This shows that the AB effect of the magnetic field always decreases the mutual information.

Notice however that the change of mutual information $I(\varphi)$ with respect to the case of zero flux diverges in the massless limit $m\rightarrow 0$ for any nonzero $\varphi$. This is because the mutual information of the mode $n=0$ diverges in one dimension for semi-infinite regions, and this is not the case for nonzero $\varphi$.

\bigskip

{\sl \bf Summary and outlook.---}
The Bohm-Aharonov effect produces changes in vacuum entanglement entropy periodic in the flux. We studied a simple example in two dimensions where it always decreases the entanglement in vacuum for non zero holonomies. This can be interpreted as a consequence of the AB interference for the modes, where the holonomy induces an effective mass for the fields. We found that the precise form of the effect is model dependent and, for particular cases, the AB oscillation of the entropy can achieve very large values.

Other scenarios where the AB can be computed are higher dimensional analogs of our calculation for free fields, amenable to dimensional reduction, or the case of a magnetic flux vortex in two dimensions using the numerical technique of Srednicki \cite{sred}. In higher dimensions one has to use mutual information in order to eliminate spurious divergences of the change on the entropy with the magnetic flux due to the change in the mass induced area terms \cite{wil}. For two regions on both sides of an annulus on a plane the variations with magnetic flux of the mutual information is not expected to diverge in the massless limit,  unlike in the example discussed in this paper. This is because this quantity is finite for the zero magnetic field case. It would also be interesting to explore this effect in the context of holography.

\bigskip

{\sl \bf Acknowledgments.---} 
This work was supported by CONICET, CNEA
and Universidad Nacional de Cuyo, Argentina.

\end{document}